\documentclass[aps,prl,noeprint,superscriptaddress,longbibliography,reprint]{revtex4-1}
\usepackage{graphicx}
\usepackage{braket}
\usepackage{amsmath}
\usepackage{amsthm}  
\usepackage{amssymb}  
\usepackage{bm}  
\usepackage[british]{babel}  
\usepackage[usenames, dvipsnames]{xcolor}  
\usepackage[hidelinks,colorlinks,linkcolor=blue,citecolor=red,urlcolor=blue]{hyperref}  
\hypersetup{hidelinks,colorlinks,linkcolor=blue,citecolor=red,urlcolor=blue} 
\graphicspath{{../figs/}}

\begin{document}

\title{Controllable finite ultra-narrow quality-factor peak in a perturbed Dirac-cone band structure of a photonic crystal slab}

\author{Alex Y.\ Song}
\affiliation{Department of Electrical Engineering, Stanford University, Stanford, California 94305, USA}
\author{Akhil Raj Kumar Kalapala}
\affiliation{Department of Electrical Engineering, University of Texas at Arlington, Arlington, Texas 76019, USA}
\author{Ricky Gibson}
\affiliation{Air Force Research Laboratory, Wright-Patterson AFB, Dayton, OH 45433, USA}
\author{Kevin James Reilly}
\affiliation{Department of Electrical and Computer Engineering, University of New Mexico, Albuquerque, NM 87131, USA}
\author{Thomas Rotter}
\affiliation{Department of Electrical and Computer Engineering, University of New Mexico, Albuquerque, NM 87131, USA}
\author{Sadhvikas Addamane}
\affiliation{Department of Electrical and Computer Engineering, University of New Mexico, Albuquerque, NM 87131, USA}
\author{Haiwen Wang}
\affiliation{Department of Applied Physics, Stanford University, Stanford, California 94305, USA}
\author{Cheng Guo}
\affiliation{Department of Applied Physics, Stanford University, Stanford, California 94305, USA}
\author{Ganesh  Balakrishnan}
\affiliation{Department of Electrical and Computer Engineering, University of New Mexico, Albuquerque, NM 87131, USA}
\author{Robert Bedford}
\affiliation{Air Force Research Laboratory, Wright-Patterson AFB, Dayton, OH 45433, USA}
\author{Weidong Zhou}
\affiliation{Department of Electrical Engineering, University of Texas at Arlington, Arlington, Texas 76019, USA}
\author{Shanhui Fan}
\email[]{shanhui@stanford.edu}
\affiliation{Department of Electrical Engineering, Stanford University, Stanford, California 94305, USA}

\date{\today}
\let\oldDelta\Delta
\renewcommand{\Delta}{\text{\scalebox{0.75}[1.0]{$\oldDelta$}}}

\begin{abstract}

We show that by using a perturbed photonic Dirac-cone, one can realize ultra-narrow and finite $Q$-factor peak in the wavevector space, with both the peak value and the width separately tunable. 
We also discuss a lower bound in the minimal viable width given a peak $Q$-value while maintaining sufficient $Q$ differentiation among modes. 
The strong angular and frequency $Q$-selection finds applications in optical devices where strong angle- and frequency-selection is needed.

\end{abstract}


\maketitle

Engineering the photonic band structure has led to significant advances in the developments of optoelectronic devices \cite{sakoda2004optical,Joannopoulos_book,barnes2003surface}.  
Most band-structure engineering focuses on the eigenfrequency of the modes as a function of the wavevector. 
Recently on the other hand, engineering the quality-factor ($Q$-factor) as a function of the wavevector has drawn increasing interest due to a number of applications such as light-trapping, two-dimensional lasing, and meta optics \cite{jin2019topologically,hsu2016bound,kodigala2017lasing,PhysRevLett.108.070401,hirose2014watt,koshelev2018asymmetric,Bermel:07,apl_high_q_fano,ha2018directional}. 
One of the desired features in such engineering is to obtain a fully-controllable \emph{finite} and \emph{narrow} high-$Q$ peak.
This allows the devices to interact with the electromagnetic waves only at a certain angle and frequency, which is important in angle-selective optical devices such as absorbers \cite{Takeda2019,Zhu2016,Zhou2016,Kosten2013,Hwang2017,Morita2017,Takeda2017,Woo2017,Shen2014,Sakr2017,Hamam2011,Piper_ACSPhoto_2014} and thermal emitters \cite{Kats2019nanophotonic,PhysRevApplied.4.014023,Wasserman_APL_2011,Sakr:17,GARIN201522,Xuechu_Shen_Nanoscale_2019,campione2016directional,greffet2002coherent,Han:10,PhysRevB.86.035316,Noda2012conversion,PhysRevB.72.075127,celanovic2008two,kosten2013highly}.
This feature is also useful in achieving single-mode large-area lasing for high-power applications. 
Such devices have always been challenging since as the device-size scales, the spacing between the modes in the wavevector space reduces, leading to a diminishing $Q$- and hence threshold-difference \cite{Chuang2009, Noda2017,631287,Yoshida2019,Inoue2019}. 

By exploring a perturbed photonic Dirac-cone \cite{huang2011dirac,zhen2015spawning,sakoda2012proof}, in this Letter we demonstrate a ultra-narrow finite $Q$ peak in a photonic-crystal slab. Both the peak $Q$-factor and its width in the wavevector space are independently controllable. 
Our approach exploits the strong mixing between the Dirac-cone bands to the linear order in the wavevector. 
Such mixing leads to a drastic $Q$ reduction away from the Brillouin zone center $\Gamma$. 
The peak $Q$ can be tuned by the strength of the perturbation, i.e. the size of the additional small holes in the photonic-crystal slab. The width of the peak can be tuned by the thickness of the slab. 
We also derive a trade-off relation between the peak $Q$ value and the width, which gives a lower bound of the latter given the former. 
Our construction can be used to fabricate perfect absorbers that are both direction- and frequency-selective. 
The design of surface-emitting lasers can also benefit from this result, as a controllable outstanding high-$Q$ mode at $\Gamma$ promotes single-mode lasing in a scaled device. 

\section{Main result}

\begin{figure}[b]
    \centering
    \includegraphics[width=\linewidth]{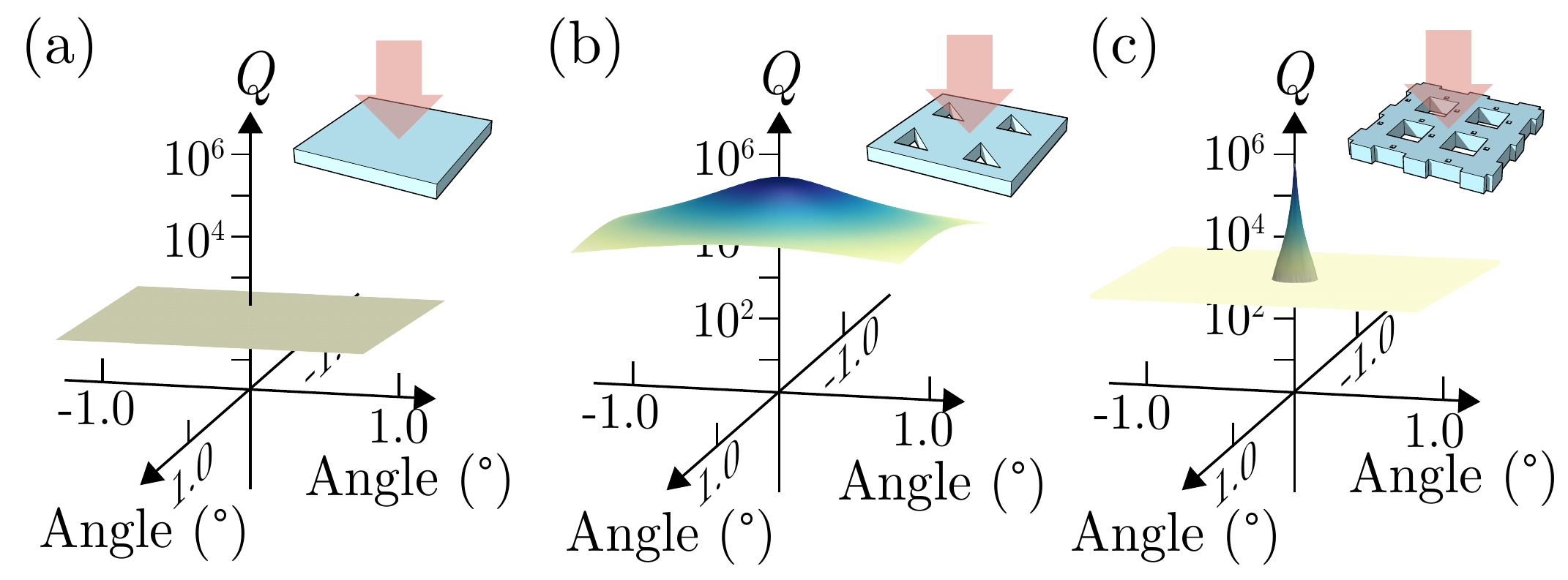}\\
    \caption{
    Comparison of the angular $Q$ variation in different structures. (a) A Fabry-P\'{e}rot band at the frequency of $0.5\, c/d$ in a uniform slab of thickness $d$ and permittivity of 12, where $c$ is the light speed. (b) A guided resonance mode at the frequency of $0.36\, c/l$ in a photonic-crystal slab, where $l$ is the periodicity of the photonic-crystal. The slab has a thickness of $0.8\,l$. The side length of the  isosceles right triangle holes is $0.5\,l$. (c) The quadrupole band for the perturbed photonic-crystal slab structure shown in Fig.~\ref{fig:cone}.
    }
    \label{fig:Qk_slab_PhC_slab}
\end{figure}

To motivate our results, we consider generically how a peaked $Q$ can be obtained. In the following, we examine optical resonators with light incident or output in the out-of-plane direction, with $\bm{k}=(k_x,k_y)$ representing the 2D in-plane wavevector, as is illustrated in Fig.~\ref{fig:Qk_slab_PhC_slab}. Our goal is to obtain a peaked $Q$ as a function of the angles.
We start with a uniform dielectric slab, which can be found in vertical-cavity surface-emitting lasers (VCSELs). There are Fabre-P\'{e}rot resonances in the vertical direction in the slab. Their $Q$-factors as a function of the wave incident angle is nearly constant near the normal direction, as is shown in Fig.~\ref{fig:Qk_slab_PhC_slab}a. 
This can be improved by introducing a 2D photonic crystal in the slab, as is illustrated in Fig.~\ref{fig:Qk_slab_PhC_slab}b.
Here, the modes at non-zero angles have a lower $Q$ than the mode at the normal direction, i.e. the Brillouin zone center $\Gamma$. 
When used in a photonic-crystal surface-emitting laser (PCSEL), the mode at $\Gamma$ would exhibit a lower threshold than other modes.
This is in accordance with the recent success of PCSELs, where single-mode lasing is maintained in a much larger area compared to the VCSELs \cite{hirose2014watt,Yoshida2019,Inoue2019}. 

Our task here is to further narrow the $Q(\bm{k})$ peak to such as the one shown in Fig.~\ref{fig:Qk_slab_PhC_slab}c. 
Our design concept is as follows. 
In a photonic-crystal slab, there exists some high-$Q$ modes and low-$Q$ modes at $\Gamma$. 
We can further assume that some modes have different symmetries at $\Gamma$, such that they won't mix and moreover it is possible to tune the structure so that the real part of the frequency of these modes are degenerate. 
At a non-zero $\bm{k}$, since the symmetry is lowered from $\Gamma$, 
these modes starts to mix with each other.
Generically, such a mixing leads to a high-$Q$ peak at $\Gamma$. 
The width of the peak is controlled by the strength of the $\bm{k}\cdot{\bm{p}}$ terms that mix the modes \cite{sakoda2012proof}.
At small $\bm{k}$, the strongest possible such term is to the linear order in $\bm{k}$.
If we consider a high-$Q$ mode $\ket{q}$ and a low-$Q$ mode $\ket{d}$ at $\Gamma$, generically, the narrowest $Q$ peak should be described by the following effective $\bm{k}\cdot\bm{p}$ Hamiltonian
\begin{equation}
    h(k) = \begin{pmatrix}
        i\gamma_d & iv_g k \\
        -iv_g k & i\gamma_q
    \end{pmatrix},
    \label{eq:h_2x2_ptb}
\end{equation}
along a direction $\bm{k}$ in the Brillouin zone. Here $k=|\bm{k}|$ is the magnitude of the in-plane wavevector, $v_g$ is the group velocity of the bands, $\gamma_d$ is the radiation constant of the low-$Q$ mode at $\Gamma$, and $\gamma_q$ is that of the high-$Q$ mode at $\Gamma$.
The resulting $Q(\bm{k})$ function for the band containing mode $\ket{q}$ is plotted in Fig.~\ref{fig:Qk_slab_PhC_slab}c, assuming the parameters of $\gamma_q=6.7\times10^{-7}$, $\gamma_d=3.8\times10^{-4}$, and $v_g=0.11\, c/2\pi$. 
Due to the strong mixing that is linear in $\bm{k}$, the $Q$ factor reduces drastically away from $\Gamma$, leading to a finite and narrow $Q$ peak.

The remaining task is to design a structure that gives the effective Hamiltonian in Eq.~(\ref{eq:h_2x2_ptb}). 
The linear mixing in Eq.~(\ref{eq:h_2x2_ptb}) corresponds to a Dirac-cone band structure \cite{huang2011dirac,zhen2015spawning}. 
In a photonic crystal slab, one can form a Dirac cone by creating an accidental degeneracy at the $\Gamma$ point between a pair of two-fold degenerate dipole states and a singly degenerate state \cite{huang2011dirac,zhen2015spawning,Minkov_zero_index_BIC}. 
However, in all previous works, due to the symmetry of the structure used, the singly degenerate state has an infinite $Q$. The resulting system thus does not have the desired $Q(\bm{k})$ dependency \cite{sakoda2012proof}. 
Our approach is to start with a Dirac-cone formed by the approach described above and perturb it such that the infinite-$Q$ mode becomes radiative.

\begin{figure}[t]
    \centering
    \includegraphics[width=\linewidth]{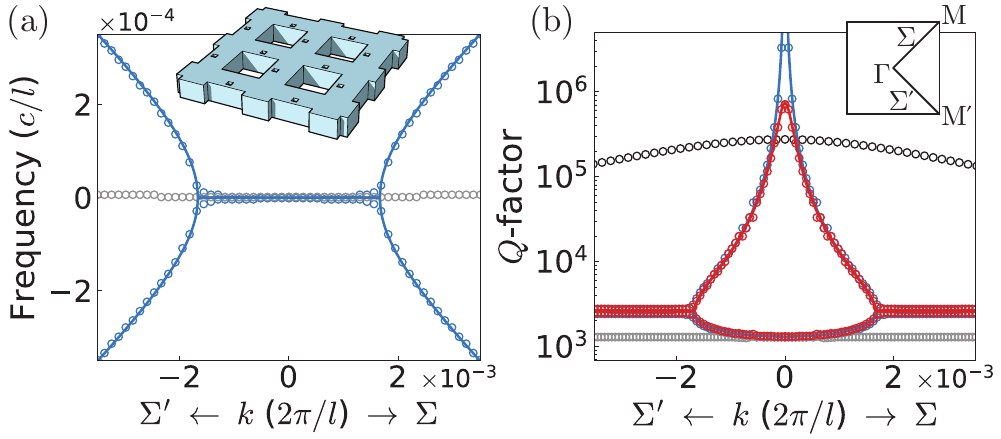}\\
    \caption{
    Band structure and the $Q(\bm{k})$ function of the photonic Dirac-cone. Circles represent numerical calculation results, while solid curves represent results of the effective model in Eq.~(\ref{eq:h_2x2_ptb}). 
    (a) band structure of the unperturbed structure. Two dispersive (blue) bands and the flat-band (gray) are plotted. Inset: schematic of the photonic-crystal slab. The frequency is offset to the Dirac-cone frequency at $0.496\,c/l$.
    (b) $Q(\bm{k})$ functions of the unperturbed structure (blue) and the perturbed (red). The flat-band is plotted in Grey circles. Black circles represent the $Q(\bm{k})$ function of the photonic-crystal slab with triangular holes. Inset: diagram of the first Brillouin zone. 
    }
    \label{fig:cone}
\end{figure}

To implement our approach, we consider the photonic-crystal slab structure as is shown in the inset of Fig.~\ref{fig:cone}a. The photonic-crystal slab has a square lattice with period $l$ in both $x$ and $y$ directions. We normalize all the other geometrical parameters in units of $l$. The slab has a thickness of $t=0.8\,l$, and a permittivity of $12$ as is typical for common semiconductors in the infrared wavelength range. In each unit-cell, there is a large square-shaped hole at the center with a side length of $a_1$ and two smaller holes with a side length of $a_2=0.05\,l$ located at $(0.4\,l, 0)$ and $(0, 0.4\,l)$, respectively. The large hole is through the slab, while the smaller holes have a depth of $t_2=0.05\,l$. Without the smaller holes, the PhC slab has a $D_{4h}$ point-group symmetry \cite{inui2012group}. 
The symmetry reduces to $C_{2v}$ with the smaller holes added. 

The band structure and the $Q(\bm{k})$ functions can be directly calculated by monitoring the poles of the scattering matrix on the complex frequency plane \cite{song2018first} \footnote{See supplemental material for more information.}.
The results are plotted as blue circles in Fig.~\ref{fig:cone}a and b, respectively.
In the absence of the smaller holes, and with a side length of $a_1 = 0.69\,l$, three modes are tuned to degeneracy at $\Gamma$. 
In the resulting band structure in Fig.~\ref{fig:cone}, two of the bands are dispersive, while the third forms a flat band.
Both the band structure in Fig.~\ref{fig:cone}a and the $Q(\bm{k})$ functions in Fig.~\ref{fig:cone}b show two phases separated by exceptional points \cite{zhen2015spawning}.
At $\Gamma$, the infinite-$Q$ mode belongs to the $B_{1g}$ representation of the $D_{4h}$ point group and is of quadrupole nature. 
The low-$Q$ mode belong to the $E_u$ representation, which is of dipolar nature and strongly couples to free-space radiation. 
The resulting $Q(\bm{k})$ is sharply peaked for one of the bands. 
However, the mode at $\Gamma$ in this band has a infinite $Q$-factor and is a bound state in the continuum (BIC) \cite{hsu2016bound}, which can not couple to free space to perform light detection, absorption, or emission.  

To create a sharply peaked $Q(\bm{k})$ where $Q$ is finite at the peak, we perturb the structure by introducing two smaller holes as mentioned above to break the $D_{4h}$ symmetry, such that the quadrupole mode is allowed to radiate. 
With the perturbation, the real part of the band structure is virtually identical to the the    one without in Fig.~\ref{fig:cone}a, hence is not plotted.
The $Q(\bm{k})$ functions with the perturbation are shown as red circles in Fig.~\ref{fig:cone}b. 
We observe that the $Q(\bm{k})$ functions for both the high-$Q$ and the low-$Q$ bands are very close to those without perturbation, except near the Brillouin zone center $\Gamma$. 
At $\Gamma$, the high-$Q$ mode now has a finite $Q$-factor rather than infinite.
Away from $\Gamma$, the $Q$-factor quickly reduces, thus forming a sharp peak at $\Gamma$.
As a reference, the $Q(\bm{k})$ function of the band at the frequency of $0.36\,c/l$  in a photonic-crystal slab with a square lattice and triangular holes is also plotted in the black curve in Fig.~\ref{fig:cone}b \cite{hirose2014watt, song2018first}.
This is a singly degenerate band with no other bands nearby in frequency, hence does not form a Dirac-cone.
A schematic of this structure can be found in Fig.~\ref{fig:Qk_slab_PhC_slab}b.
The peak $Q$-factor is at a level of $10^5$, similar to our design here.
However, The $Q(\bm{k})$ function is much more slowly varying. 
The $Q$ peak width is reduced by an order of magnitude in our work through exploiting the strong linear-order mixing.

The band structure and the $Q(\bm{k})$ function of the two dispersive bands are described by the effective Hamiltonian in Eq.~(\ref{eq:h_2x2_ptb}) \cite{zhen2015spawning}.
Here, $\gamma_d$ and $\gamma_q$ can be read out from the numerical calculations at the Brillouin zone center $\Gamma$. $v_g$ can be obtained by fitting to the slope of the bands outside of the exceptional points.
The complex eigenfrequencies of the Dirac-cone bands are then obtained by solving Eq.~(\ref{eq:h_2x2_ptb}).
The resulting band structure and the $Q(\bm{k})$ functions are plotted as solid curves in Fig.~\ref{fig:cone}.
The results very accurately reproduce the numerical simulations. 
In existing studies of the photonic Dirac-cones, $\gamma_q$ is zero, leading to an infinite $Q$-factor.  
In our work, $\gamma_q$ becomes finite due to the introduction of the smaller holes.
The perturbation can also cause small corrections to other parameters in Eq.~\ref{eq:h_2x2_ptb}, which are omitted in the lowest order. 
From Eq.~(\ref{eq:h_2x2_ptb}) it is clear that the off-diagonal terms that is linear in $\bm{k}$ causes a strong mixing between the dipolar and the quadrupole modes, resulting in a sharp $Q$ peak.
It is important to note that the entire band structure and the $Q(\bm{k})$ functions are controlled by only three parameters, i.e. $\gamma_d$, $\gamma_q$, and $v_g$.

\section{Independent engineering of the peak $Q$ and its width}
\begin{figure}[t]
    \centering
    \includegraphics[width=\linewidth]{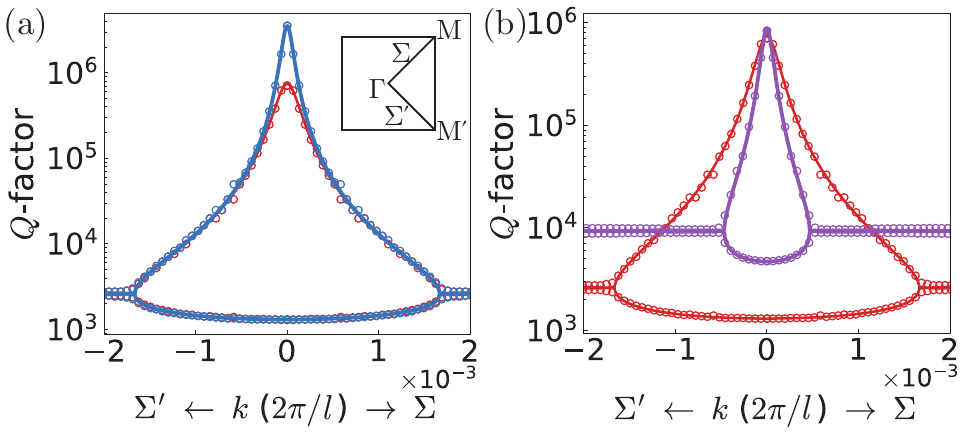}\\
    \caption{
    (a) The $Q(\bm{k})$ functions of the photonic-crystal slab with the side length and depth of the small holes to be $0.05\,l$ (red) and $0.04\,l$ (blue), respectively. 
    (b) The $Q(\bm{k})$ functions of the photonic-crystal slab with a thickness of $0.8\,l$ (red) and $0.775\,l$ (purple), respectively.
    The circles are results of the numerical calculation. The solid curves are results of the effective model in Eq.~(\ref{eq:h_2x2_ptb}).
    }
    \label{fig:tune_Q_bandwidth}
\end{figure}

In our approach to engineer the $Q(\bm{k})$ function, both the peak $Q$ and its width can be tuned. 
The peak $Q$ value can be straightforwardly engineered by controlling the size of the small holes. For example, In Fig.~\ref{fig:tune_Q_bandwidth}(a) we plot the $Q(\bm{k})$ function of the same structure as above, but with the small holes having a reduced side length and depth of $a_2=t_2=0.04\,l$. The resulting $Q$-factor at $\Gamma$ now increases by an order of magnitude while the width of the peak remains largely unchanged.

To tune the width of the $Q$ peak without changing the peak value, we can change the $Q$-factor of the dipolar mode at $\Gamma$. 
As is discussed above, the $Q(\bm{k})$ functions of both bands are controlled by only three parameters, i.e. $\gamma_q$, $\gamma_d$, and $v_g$.
By increasing the $Q$-factor of the dipolar mode, the region inside the exceptional points shrinks, leading to a reduced peak width. 
A demonstration is shown in Fig.~\ref{fig:tune_Q_bandwidth}(b), in which we compare the $Q(\bm{k})$ function of the design above that corresponds to the red  curve in Fig.~\ref{fig:tune_Q_bandwidth}a and one with a reduced slab thickness of $t=0.775\,l$ (purple). 
The side length of the large hole is adjusted to $0.695\,l$ to maintain the degeneracy of the real part of the bands. 
We also tune the smaller holes to have a side length and depth of $a_2=t_2=0.035\,l$, to maintain the peak $Q$-value at $\Gamma$.
The resulting $Q(\bm{k})$ functions of both numerical calculation (circles) and the analytical model (solid curves) are shown. 
The $Q$-factor of the dipolar mode increased by four times in the thinner photonic-crystal slab. 
As a result, the high-$Q$ bandwidth reduces by a factor of 2 while the peak $Q$ value is unchanged, as is observed in Fig.~\ref{fig:tune_Q_bandwidth}(b).

\section{Peak-width relation}
The above discussion of the tuning of the high-$Q$ width gives a hint of a relation between the minimal viable width and the peak $Q$. 
In real devices, it is often desirable to design a certain $Q$-factor at the zone center $\Gamma$ and at the same time a strong $Q$-differentiation against other modes away from $\Gamma$.
For simplicity, We define the width of the high-$Q$ region by the width of the wavevector region inside the exceptional points. 
The location of the exceptional point can be calculated from Eq.~(\ref{eq:h_2x2_ptb}) as $k_e = (\gamma_d-\gamma_q)/2v_g$.
It is clear that a higher $Q_d$, i.e. a smaller $\gamma_d$, leads to a smaller $k_e$. 
By the requirement of $Q$-differentiation, we consider that the peak $Q$ value at $\Gamma$ is at least twice the value outside the exceptional points. 
The minimal $\gamma_d$ is then given by $\gamma_d = 3\gamma_q$.
This leads to a minimal width as
\begin{equation}
    w_{\textrm{min}}=2k_e = 2\frac{\omega}{Q_{\textrm{max}}\cdot v_g}
    \label{eq:peak_width}
\end{equation}
where $\omega$ is the frequency of the Dirac point, $Q_{max}$ is the desired peak quality factor at $\Gamma$, and $v_g$ is the group velocity of the Dirac-cone bands.  
For a target $Q$-factor at $\Gamma$, Eq.~(\ref{eq:peak_width}) thus gives the lower bound on the minimal peak width while maintaining sufficient $Q$ differential from the $\Gamma$ point to other wave vectors.

\section{Angle- and frequency-selective perfect absorption}

\begin{figure}[t]
    \centering
    \includegraphics[width=\linewidth]{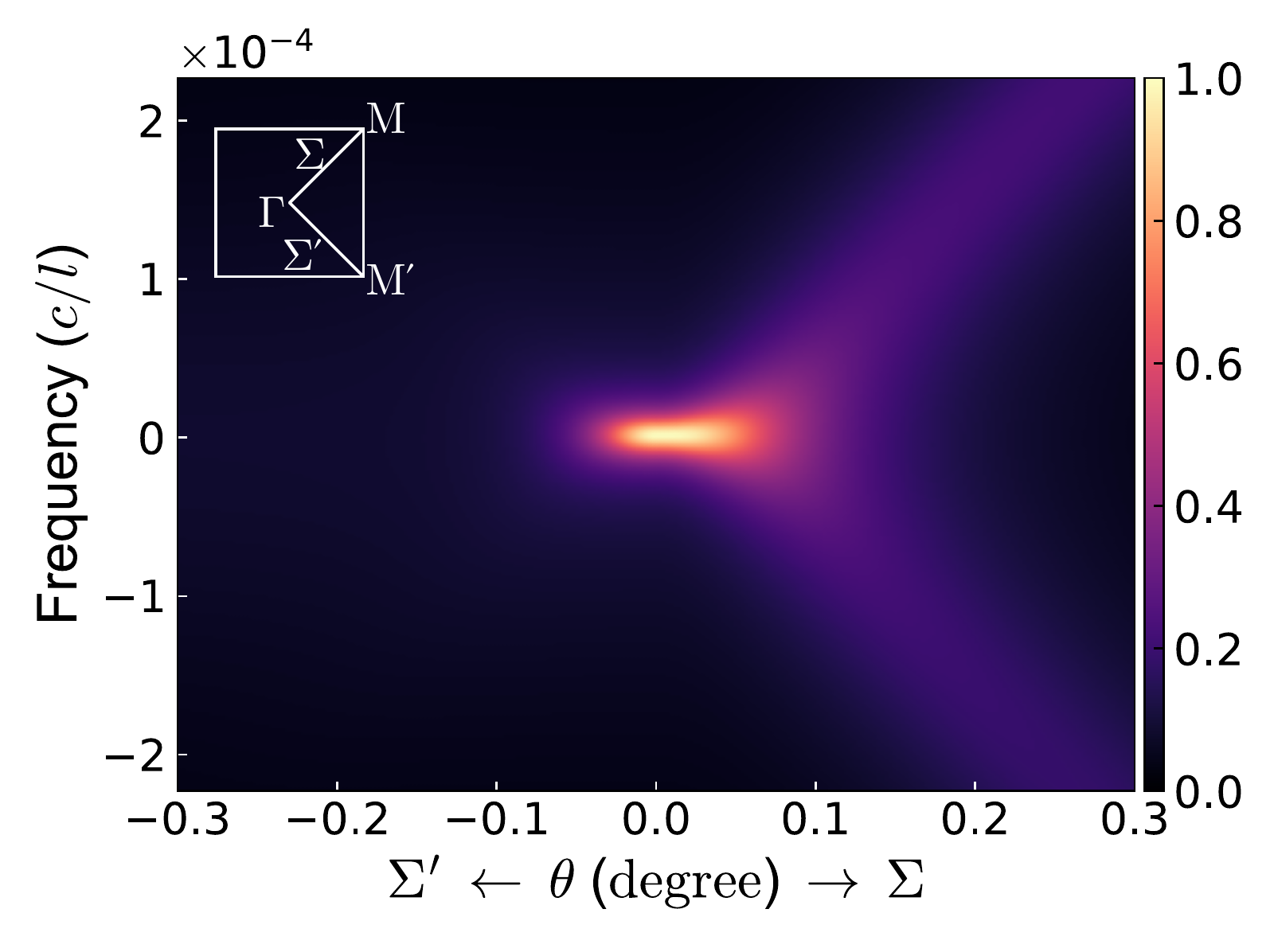}\\
    \caption{
    The absorption spectra of the angle- and frequency-selective absorber. The frequency is offset to the Dirac-cone point at $0.485\,(c/l)$. The incident wave is polarized such that the in-plane projection of the electric field is pointed in the $45^\circ$ direction.
    }
    \label{fig:absorber}
\end{figure}

A sharply peaked $Q(\bm{k})$ function is useful in a variety of scenarios where strong angle- and frequency-differentiation is needed, such as angle-selective sensors or large-area single-mode lasers. 
As an example, here we demonstrate a perfect absorber based on the photonic-crystal slab with a strong selection in both the frequency and the incident angle. 
The geometric parameters of the slab is as follows. The slab has a thickness of $t=0.85\,l$, the large hole has a side length of $0.682\,l$, and the small holes have a depth and a side length of $0.1\,l$. 
We add a mirror to the bottom of the photonic-crystal slab at a distance of $20.3\,l$. Light incidence is from the top.
We make use of the quadrupole mode at $\Gamma$, which is an outstanding high-$Q$ point in a parameter space of three dimensions, i.e. frequency plus 2D solid angles. 
We assume a uniform material loss in the photonic-crystal slab by adding a constant imaginary part to its refractive index, i.e. $\varepsilon=12-4.8\times 10^{-5}i$, representing a semiconductor interband absorption which is typically much broader than the optical resonance. 
Such a material absorption provides a modal internal loss to the quadrupole mode at $\Gamma$ in the same amount as the radiative loss, and hence resulting in a perfect absorption due to critical coupling \cite{Haus1984,WonjooSuh2004}. 
At any other angle or frequency, such condition is not met and the absorption is low. 
The numerical results are shown in Fig.~\ref{fig:absorber}, where we incident light on to the structure with varying frequencies and angles. Here, light is assumed to be polarized such that the in-plane projection of the electric field is pointed in the $45^\circ$ direction, i.e. it is $p$-polarized for the spectra along $\Sigma$ and $s$-polarized along $\Sigma'$. 
It is observed that strong absorption only occurs within an angle of $0.1^\circ$ from normal, in a narrow frequency range of $10^{-5}\,c/l$.
Again, the width of the angle can be tuned by changing the $Q$-factor of the dipolar mode, for example by tuning the slab thickness. 
We note that the adding of the bottom mirror can introduce a correction to the frequency and the radiative constant of both the dipolar and the quadrupole bands \cite{Xu2000}. Yet, this would not affect the narrowing of the $Q$ peak in the perturbed Dirac-cone.

\section{Conclusion}
In summary, in this Letter we presented an approach to obtain a finte and ultra-narrow $Q$ peak, with the peak value and width independently tunable. This is realized by employing a perturbed photonic Dirac-cone. Utilizing the linear mixing between modes in the Dirac-cone, the modes at $\Gamma$ strongly remix away from $\Gamma$, leading to a rapidly decreasing $Q(\bm{k})$ function for a high-$Q$ band. The narrow $Q$ peak is useful in scenarios where angular selection is needed, such as directional absorber or emitter. The feature can be useful in surface-emitting lasers as well, since a large $Q$ differentiation promotes single-mode lasing. 
Although we focus on 2D PhC structure, this approach applies to other dimensions such as 1D distributed Bragg reflectors (DBRs).

\begin{acknowledgments}
This work is supported in part by the Department of Defense Joint Directed Energy Transition Office (DE-JTO) under Grant No. N00014-17-1-2557. 
\end{acknowledgments}

\bibliography{library.bib}

\end{document}